# BEYOND HARM:

## AN ETHICAL FRAMEWORK  TO TACKLE MISINFORMATION ON SOCIAL MEDIA

- Draft please do not cite – (Under review) – Marianna B. Ganapini

This paper aims at building an actionable framework for permissible online content moderation to combat misinformation. Often strong content moderation policies are invoked when misinformation causes harm. By adopting J.S. Mill's ethical framework, I show the complexities involved in permissible content moderation. The conclusion will be that, besides invoking the notion of harm, we should also introduce the idea of 'cognitive autonomy' and adopt useful tools, such as cognitive nudging, to promote a healthier epistemic environment online.

### 0. Introduction

The main topic of this paper is misinformation. Misinformation is an umbrella term that I will use to refer to all kinds of problematic information, intended as "inaccurate, misleading, inappropriately attributed, or altogether fabricated" textual information (Jack 2017). This is a very broad definition of misinformation as it applies to many different things, such as memes, fake news, conspiracy theories and so on. It also applies to both false or misleading information that is shared to deceive, manipulate or confuse one's audience, and to any inaccurate information shared with no intention to deceive. I will only focus on text-based misinformation, I will not tackle video or audio deepfake (which presents its own specific set of challenges; Fallis 2020, Floridi 2018, Rini and Cohen 2021).

This paper will focus only on the potential harm of misinformation that is spread online and on social media in particular.[1] Recently, the spread of misinformation online has brought many to worry about its effects. The general worry is that widespread false and misleading news and stories online may influence people's beliefs and choices. In some cases, these choices are perceived as potentially harmful, either for those who make them or for society at large (Howard, 2021; Zaibert 2021). These harmful effects are often related to "post truth scenarios" where misinformation is a threat to democracy or to our democratic institutions. For instance, the World Health Organization has recently stated that the "2019-nCoV outbreak and response has been accompanied by a massive 'infodemic'" and has pledged to be "working closely to track and respond to myths and rumours" (WHO 2020).

As a result of some strong public pressure, social media companies have decided to limit the spread of fake news and the like by moderating the content circulating on their platforms. Content is curated manually, automatically (using AI) or through a users' flagging system. When platform rules are violated, users may face considerable repercussions: their accounts may be closed and they may be banned from posting content for months, with little hope for appealing the platform's decision. For clarity, in this paper I will divide up content moderation in the following way:

---

[1] It is beyond the scope of the paper to weigh on the harmful impact of cyberbullying and child pornography.

**Strong Content Moderation:** removing posts, blocking users, preventing users from posting certain content, substantially down-ranking specific content and hashtags. This type of content moderation is directed to tackle specific content and to limit its spread.

**Moderate Content Moderation:** boosting content, adding warning labels, counter-speech, fact-checking. This type of content moderation tackles specific content but does not directly limit its spread.

**Light Content Moderation:** nudging & slugging, friction, inoculation strategies. This type of content moderation is not directed to any specific content but addresses structural issues.

I will mostly focus on the most stringent form of content moderation because it has raised a lot of controversy. I will have very little to say about moderate forms of content moderation (Lepoutre 2021), but at the end of the paper I will argue for the need of adopting light content moderation strategies.

To counter misinformation about the Coronavirus, Twitter has promised to ban potentially "harmful" content that "promotes unverified claims and misinformation about coronavirus". Similarly, they pledge to remove content that "has a clear call to action that could directly pose a risk to people's health or well-being" (Gadde and Derella 2020). Facebook and other platforms have made similar pledges, all based on the rationale that some content, when widespread, can cause harm.

These attempts have been met with some considerable resistance by those who think that such strong forms of content moderation often constitutes a violation of our freedom of speech (Mudde 2020; The Economist 2020; United Nations 2021). As a right, freedom of speech cannot be exercised against social media platforms because these are private companies. Similarly, these companies do not fall under the First Amendment of the US Constitution. That said, it seems reasonable to expect them to be committed to upholding human rights and free speech in particular (Simpson, 2020: 290). As a result, the discussion is now centered on whether harmful misinformation should be banned from social media and how and when platforms should intervene.

My paper is an attempt to answer these questions and to introduce best practices for deliberating about strong content moderation.[2] To help me, I will adopt some of the machinery introduced by J.S. Mill in his analysis of liberties and freedom of action and discussion. To be sure, this is not meant to be an exegetical paper about Mill's philosophy and, though I will adopt some of the key insights of his ethical framework (e.g., Harm Principle), I do not expect this to be a Millian response to the problem of misinformation. Also, I will not tie my discussion to a particular legislative framework, and the goal here is not to establish what laws and regulations should be set in place to counter misinformation.

Here is how I structure the paper. First, I will introduce the idea of harm and Mill's Harm Principle. I lay out three necessary conditions for the Harm principle to apply to misinformation and show that they are quite difficult to meet. Second, I will look at some key elements platforms need to take into account before enforcing severe content moderation policies against harmful misinformation (again here I take my lead from Mill). Third, I will look at one specific use-case. Four, I will argue that platforms are morally required to adopt

---

[2] This way to proceed is not new in the field of ethics of technology and ethics of AI in particular where norms and guidelines have been offered to tech companies, governmental institutions and the public at large. See Floridi, et al. (2018) and European Commission (2021).

*light* content moderation policies even against non-harmful misinformation and protected speech, and explain what those policies could be.

## 1. When Does the Harm Principle Apply?

In this section I lay out several conditions for the Harm principle to apply to false speech. I also assume that, when the Harm principle applies, speech loses protection.[3] I start with the notion of harm coming from J.S. Mill's *On Liberty*: a harmful action undermines people's fundamental interests and rights. This is a fairly strong notion of harm that requires that one is harmed if their vital interests (i.e. interest for which they have a right) are infringed upon by someone's action. In this context, Mill distinguished harm from mere offense (though see *On Liberty*, V 7). For instance, just being emotionally hurt or offended by someone's words or actions is not enough for those words and actions to count as harmful. This notion of harm has been met with some resistance.[4] However, for the purpose of this paper we can leave it undetermined how strictly or openly we define harm while knowing that 'harm' constitutes the basis that allows for limitations of one's liberties, according to Mill's Harm Principle:

---

[3] I remain open to the possibility that harm may not be a sufficient condition for speech to lose protection as other conditions may be needed (e.g. imminent harm). Since this topic is already widely debated in philosophy, I cannot take it up here and will only mention it in passing when relevant to my analysis.

[4] Some complain that Mill's principle is overly restrictive (Feinberg, 1984), and argue that attacks to one's emotions, health and autonomy should also be part of what constitutes harm.

"That the only purpose for which power can be rightfully exercised over any member of a civilised community, against his will, is to prevent harm to others" (Mill 1On Liberty, ch 28).

Thus, the Harm Principle seems to work as a preventive measure to avoid harm (cf O'Rourke 2001: 154): it can be rightfully applied in those cases (and only in those cases) in which an action leads or is expected to lead to some harm. For Mill, the Principle does not only regulate governments' actions and legislations, but it applies to our society at large.

Let's now see how this would apply to misinformation. As we mentioned, there is a tendency to think that some misinformation is harmful and needs to be banned or blocked (The Lancet, 2020). Though this kind of considerations do not necessarily directly invoke Mill's Harm Principle, they tend to refer to the idea that, if misinformation endangers others, then it should not be protected. As Howard (2021) puts it, we have a duty not to endanger others and when the fake posts we see online are in fact dangerous, we have a duty not to endorse, post or retweet them.[5]

I see three elements of contention for this kind of analysis:

a. It is impermissible to restrict free speech online if the speech in question is not an actual cause of harm (but it is simply correlated with the occurrence of harm). This is not an idle point when it comes to misinformation as we need to be careful about the causal power attributed to it. Take vaccine hesitancy. Many believe that misinformation about vaccines drives people's choices about whether to get vaccinated. And yet these claims are often made with little evidence. Contrary to that, misinformation about vaccines may not be the real cause of vaccine hesitancy. Vaccine hesitancy may correlate with being exposed to misinformation because they might be both caused by some other third factor (e.g. pre-established antivax sentiments, sense of loss of control, distrust in the government, political affiliation). Here's one example: being a Trump supporter may cause both vaccine hesitancy *and* an interest in antivax websites and social media groups. Hence, it is possible that being exposed to misinformation is not what is playing a key role in the decision to avoid vaccines: it is the political affiliation to a certain group (and the desire to fit in) that may well be the real culprit (Mercier 2020). Hence, because of these confounding factors, there is often no straightforward causal path from widespread online misinformation to harmful actions. To boot: tough widespread in certain circles and communities, the amount of overall misinformation online is not as much as many researchers tend to believe (Mercier 2017; Altay & Acerbi 2022). People's news diet is generally not riddled with fake posts, and only those who are already fairly radicalized consume large amounts of fake news. This point invites caution in attributing any real cause power to misinformation online and more resources should be put in studying these trends (see Haidt and Bail ongoing).

---

[5] Zaibert (2021) also links misinformation and harm.

b. It is impermissible to restrict free speech online if the probability of harm being caused is low. Even assuming that misinformation causes harm, to apply the Principle we still need to assess how probable that harm actually is. High probability may warrant coercion and the application of the Harm Principle, whereas low probability might not justify limiting freedom of speech. Again, assessing probabilities is no small feat when it comes to misinformation online. Often online harm is caused as the result of the aggregate of low-impact, harmless acts. Single posts, which make some falsehoods viral, have limited effects in themselves. Thus here we face a version of the so-called "many hands problem" (Thompson 1980; Nissenbaum 1996; Sinnott-Armstrong 2005; Bovens 2010): single falsehoods can hardly cause any harm, it is the aggregation of them that may bring about harmful effects. Thus, it is unclear whether single false posts - whose causal efficacy is minimal - would justifiably be stripped of protection (Hart and Honoré 1959).

c.  It is impermissible to restrict free speech online if the harm produced is justified in some way (Howard, 2021). Even if some speech is clearly harmful, there might be reasons for protecting it. In general, it may be reasonable to allow harmful acts if these are part of an equitable system in which everyone is taking some risks because they also enjoy some fairly distributed or compensatory benefits as a result. Consider this: we happily drive children around in our cars even though driving is a fairly risky activity that can cause serious harm to them. We do so because we recognize that the benefits they enjoy from going places outweigh the risks. Similarly, one might argue that, even if some false speech is harmful, a system that allows harmful false speech to exist and even proliferate on social media is *equitable* because it still allows everyone to enjoy the benefits of sharing their opinion while potentially producing harmful speech as well. Hence, when analyzing free speech in relation to misinformation, we need to determine first whether any produced harm is part of an equitable risk distribution system. If so then, even when dangerous, misinformation may enjoy (some degree of) protection.

In conclusion, establishing whether the Harm Principle applies in banning misinformation depends on both empirical and theoretical matters. Any analysis or deliberation that calls for strong content moderation policies should show that *at least* the issues mentioned above have been addressed.

## 2. Three Restrictions on (Strong) Content Moderation

Even once we have established when and how misinformation falls under the Harm Principle, we still have to determine what social media platforms should or may do. This point can be explained by once again looking at Mill who maintained that applying the Harm Principle is a necessary (but not sufficient) condition to ban harmful speech. That is, Mill believes that coercion needs to be justified by a Utilitarian deliberation process:

> "[I]t must by no means be supposed, because damage, or probability of damage, to the interests of others, can alone justify the interference of society." (Mill, On Liberty, V 3)

Mill seems to indicate that even if an act is harmful that is not sufficient reason for coercion. The Principle provides *some* (pro-tanto) reasons for intervening. However, that is not enough to *all things considered* justify those restrictions. Thus, depending on the situation, social media platforms may be forbidden, permitted or mandated to enforce strong content moderation policies to curb harmful misinformation. And here are three considerations they need to take into account when deliberating about this:

**One: overall consequences.** The fact that some misinformation is clearly harmful does not mean platforms should automatically enforce content moderation policies. Companies may realize that stifling a certain type of content may lead to greater harm in some way (Mill, On Liberty, IV 3). In other words, general welfare needs to be taken into consideration when determining permissible content moderation. For instance, a social media company may realize that blocking some users - whose tweets may qualify as harmful - may prevent them from voicing dissent against their oppressive government. Thus, even if the harm produced is not justified, any strong corrective action may produce even worse effects.

**Two: imminent and inevitable harm**. Platforms may stop misinformation if it produces harm that is imminent or inevitable. It is indubitable that some falsehoods have a *high probability* of generating harm. And at times the harm caused is serious and cannot be prevented. Here is a classic example:

> "When falsehoods are banned, it is not only because they are falsehoods, but also because they threaten to produce real harm. Consider the canonical example of an unprotected falsehood: a false cry of "fire!" in a crowded theater. Such a cry is not merely false. It also threatens, with a sufficiently high probability, to cause serious

harm — and under plausible assumptions, little or nothing can be done, in time, to prevent that harm." (Sunstein, 2020, p 405)

Falsely crying 'fire!' in a crowded hall is the kind of act that has a high probability of causing serious harm while leaving little time to prevent harmful consequences from occurring. The harm in this case is considered to be *imminent*. This is the type of speech that is usually not legally protected and there seems to be an agreement that this type of act should be banned.[6] There are more controversial cases: these are the cases in which the harm is not imminent yet very likely to happen and difficult to prevent. Inciting violence, fomenting racial hatred, calling for acts of terrorism are all types of speech that are often dangerous in so far as they are likely to cause harm and such a harm may be particularly difficult to stop from happening.[7] Falsehoods online may also constitute a form of dangerous speech too. In Myanmar, for instance, lies spread on Facebook were key to the violence that exploded against the Muslim Rohingya minority group in 2018. This is not a rare phenomenon: hostile rumors are often used in ethnic riots as a (disguised) call for action (Horowitz, 2001). On this view, rumors spreading are a means to coordinate among rioters and actually start the attack. And this violence is often hard to stop in its tracks. Arguably, it should be required that platforms stop the spread of misinformation in this case too (Howard, 2021).

---

[6] Similarly, defamation, perjury, false advertisement about cures and drugs and falsely presenting oneself as police officer, are all acts that are likely to produce serious harm and are not protected (Sunstein, 2020).

[7] There are countries, the US for instance, where this speech is legally protected. Also, a good chunk of the philosophical tradition claims that this speech deserves protection even if it has a high probability to generate harm (Scanlon 1972).

Finally, if the harm misinformation is likely to cause is neither imminent nor inevitable, platforms may *not* be required to enforce strong content moderation policies (Feinberg, 1984, 1985). Platforms may be able to diffuse the harm by adopting other types of policies, such as counter-speech or other moderate forms of content moderation (e.g. fact-checking).

**Three: direct vs indirect cause of harm.** When deliberating about content moderation, we also need to consider whether it is possible to intervene on the direct causes of the harm, rather than focusing on the indirect ones which may bear less causal responsibility for the harm.

Let's first zoom in on the distinction itself. A directly harmful act is an act that *itself* causes some harm. When successful, murder or violence more generally are acts that have a direct impact on people's well-being. The key feature of direct harmful acts is that the very goal of the action turns out to be something harmful (even if the agent did not intend to do anything harmful). Failing to vaccinate your kids is a direct harmful act (for them and those around them and even if your intention were to do good). Indirect harmful acts are part of the causal chain that may lead to a dangerous situation/event which directly produces harm. Because they are part of this causal chain, they could be seen as causes of harm but the harmful event is a byproduct rather than the goal of the act itself. Thus, other things with equal intervening on direct causes seems more reasonable than applying coercion to indirect causes (Alexander 2000).

Speech can be harmful either directly or indirectly. Examples of false speech that is a *direct* cause of harm are defamation and, arguably, hate speech. Defamation brings harm because it constitutes an attack to the reputation of someone who has done nothing wrong. Similarly (though more controversially), some have argued that it is constitutive of hate speech that it produces harm by depriving others of their dignity. As Waldron puts it, dignity "is a matter of status—one's status as a member of society in good standing—and it generates demands for recognition and for treatment that accords with that status." (2012, p 60) Waldron indicates that engaging in hate speech is tantamount to depriving some people of the recognition demanded by their dignity. If so then engaging in the act itself brings about harm in a direct way and so it may be argued that platforms should stop the spread of that kind of speech.

In contrast, some misinformation only harms indirectly. To illustrate, misinformation about vaccines *may* cause harm by pushing people to *act* in a harmful way (and it is thus an indirect cause of harm). That means that, even if we were to establish that anti-vax propaganda was a or the leading cause of vaccine hesitancy, platforms may still have reasons for *not* enforcing strong content moderation on that kind of speech: vaccine hesitancy may be curbed by requiring vaccination (as it happened for Covid-19 in some countries), and if so there might be no requirement for platforms to curb the spread of anti-vax misinformation itself. That means that indirectly harmful misinformation puts less burden on the shoulders of social media platforms than directly harmful misinformation: the letter but not the former usually requires them to intervene by using strong content moderation.

## 3. A Use Case

Here I will discuss a practical example of how one might adopt the insights mentioned above to deliberate about strong content moderation. It's July 28, 2020: Twitter removed a fake post re-shared by Trump falsely claiming we finally had a cure for COVID-19. They worried that such a false statement would indirectly cause harm by, for instance, making citizens less compliant with mask mandates.[8]

As mentioned above, one of the key issues to be addressed is the causal relation between misinformation and harm, as we need to ask if some misinformation really causes any harm. In this case of the Covid cure tweet, the harmful effect produced would be brought about by, e.g., people avoiding masks. Hence, here the question is whether misinformation about covid could indirectly push people to stop wearing masks. There are two ways in which this could happen. First, people could be persuaded by the content of the misinforming tweet and feel they do not need masks anymore. Second, people could understand this tweet as signaling a way to show allegiance to Trump. On this view, this bit of misinformation actually signals the kind of behavior (i.e. no masks!) that Trump supporters should adopt, and has nothing to do with the content of the misinformation persuading anyone about the existence of a cure for covid. No matter which one of these two hypotheses is correct, the point is that, according to both, sharing this type of content might lead to harmful behavior. It is obviously an empirical matter whether this is the case or not, and we should demand of social media companies that they inform their content moderation decisions by looking at the relevant evidence.

Moreover, when the probability of harm being caused is high, then speech loses its protection. In this particular case, the fact that the tweet was shared by Trump likely increased the probability that harm would take place, even if the post had been retweeted many other times already. This addresses the "many hands" objection mentioned earlier: in this case we can pin down a single source as the *key* factor in the causal chain leading up to the potential harm. The probability of the harm taking place may vary, but given that in this instance the source (the President's Twitter account) was visible to many more people than the average user's account, that suggests that the re-tweeted (by Trump) tweet could be rightly considered 'dangerous'. Moreover, because of its role, the President enjoys more credibility and has more power to sway people's minds. As such, some of the misinformation spread from the President's account is bound to have a greater impact.

---

[8] https://www.nbcnews.com/politics/donald-trump/twitter-removes-tweet-highlighted-trump-falsely-claiming-covid-cure-n1235075

A third factor to consider is the notion of "equitable risk": the risk of allowing some politicians' falsehoods to spread may be reasonable only insofar as there are enough compensatory benefits gained by the general public (Hansson 2003). In contrast, it may be permissible to restrict speech when the risks (for the public) generated by some misinformation are greater than the benefits. In the case of Trump's retweet, there were no clear compensatory benefits in letting it spread, and Trump had been constantly and freely tweeting harmful misinformation in the past.

In conclusion, one may ask, was Twitter's decision-strategy reasonable? They decided to not only eliminate Trump's retweet but also the original post about Covid's cures. Given the severity of the situation and the potential harm to the public, a misleading tweet that gains such "credibility" online is quite risky. Because of the amplification and visibility granted by Trump's retweet, it was reasonable to expect that the impact of the original tweet itself had increased dramatically by then, and with that also its potential damage. Though such damage was neither imminent nor inevitable as mask mandates and other restrictive measures were enforced in many places, it could be reasonably argued that this type of misinformation could make it more difficult for authorities to impose those reactive measures. If this is true, then it may be possible to conclude that Twitter's decision to eliminate the tweet altogether was justified.

### 4. Beyond Harm: Cognitive Autonomy

When we talk about misinformation and harm, we often mean to convey the idea that it is the sheer *quantity* of misinformation that is detrimental to our values, wellbeing and even democratic institutions (Zaibert 2021). It is not this or that particular bit of misinformation that may cause problems: it is misinformation online that is the issue. And yet the last thing we want is for social media companies to shut down all the posts they deem fake or misleading. We do not want them to be arbiters of truth, for one thing, nor do we want to eliminate all false information, even if we do acknowledge that the constant rising volume of online misinformation – especially in certain circles and communities – may be a worrying trend.[9]

So what options do we have to curb online misinformation? I believe we should adopt a different tool in our ethical toolkit, in a way that complements the notion of 'harm'. Even when not directly and straightforwardly harmful, misinformation could be seen as a sign of a *dysfunctional and polluted epistemic environment*, in which it is difficult to exercise our autonomy. Mill defines autonomy as "the capacity to be one's own person, to live one's life according to reasons and motives that are taken as one's own and not the product of manipulative or distorting external forces, to be in this way independent" (XXX).  Building on this, I use here the notion of "cognitive autonomy" as the ability to think and direct one's reasoning according to reasons that are not the product of manipulative or distorting external forces but the result of employing our rationality and good epistemic processes. My focus is thus not on action but on the *autonomy of reasoning* (Scanlon 1972).

---

[9] As mentioned before, this worry may be overblown (Mercier 2020).

Social media is responsible for diminishing our cognitive autonomy. This is a causal claim for which we have evidence: because of the structure of our online information landscape, misinformation spreads by taking advantage of some human irrational cognitive habits. Because of how information is structured online, misinformation bypasses our capacity to reason and deliberate. As such, our online information landscape likely undermines our autonomy as cognitive agents which opens the door to misinformation. Here the claim is not that misinformation causes a loss of autonomy: it is the structure of social media that makes it difficult for us to exercise our autonomy when forming and sharing beliefs and opinions. The claim is that social media content is packaged and delivered in such a way that it makes it more difficult for us to reflectively screen it. On this view, misinformation online is thus one of the symptoms of a problematic epistemic environment that hampers our autonomy.

There are several cognitive habits which seem to result in making us vulnerable to external, manipulative epistemic forces. In this paper I will focus on one that is well documented in the psychological literature: lazy and emotional thinking (Pennycook, Rand 2019).

Humans are at times lazy reasoners, they tend to use unconscious heuristics and biases to think and make decisions rather than employ more reflective reasoning. There is evidence that things are even worse on social media where the rules of interaction are fuzzy, the point of sharing information is poorly defined and the entire structure rewards fast but shallow interactions Pennycook & Rand, 2021). If so then the type of reasoning we employ on social media is likely to be driven by low-level cognitive processes. Nguyen (2021) provides a good analysis of this phenomenon. He explains that Twitter, for instance, adopts a gamification-structure that incentivizes the 'wrong' communicative goals and norms: "Twitter gamifies communication by offering immediate, vivid, and quantified evaluations of one's conversational success. Twitter offers us points for discourse; it scores our communication. And these game-like features are responsible for much of Twitter's psychological wallop. Twitter is addictive, in part, because it feels so good to watch those numbers go up and up." Searching for and sharing true information is not really the goal of social media engagement. If Nguyen is right, what we care about while on social media is being emotionally rewarded. Divisive, controversial and ludicrous content is often more rewarding than sharing well-reasoned truths (Acerbi 2019). It follows that the current structure of social media is not conducive to engaging in reflective reasoning and exercising our rational skills.

To be sure, we harbor more cognitive habits than just lazy and emotional thinking.[10] However, the analysis above is enough for making the claim that content moderation should focus on the structural elements of how platforms are currently built: it's less about the harm caused by specific bits of misinformation and more about the large-scale, dysfunctional epistemic environment these platforms are creating that surreptitiously pushes us to be less and less autonomous in our epistemic practices online.

---

[10] Another well-documented cognitive weakness is motivated reasoning. In a polarized environment we may be prone to accept information aligned with our values and group-motivations. At times, when asked to reflect more on why and what they think, we dig their heels in and confabulate instead of adopting a critical stance. Misinformation can then be seen as also the result of this confabulatory process of finding reasons that justify one's positions or sentiments. Social media platforms may be responsible for creating epistemically problematic environments where users are incentivized to use reasoning in a confabulatory manner (to reinforce pre-seeded opinions).

## 4.1. Nudging for Cognitive Autonomy (light content moderation)

In this final section, I will point to one specific set of tools that could be used to (partially) address the structural problems of social media. Numerous studies in social and behavioral psychology indicate that if people are prompted to engage in careful reasoning, they become more alert to fake news (van der Linden et al 2017). We can call the tools that push us to become more careful reasoners 'cognitive nudges'. In the vast literature on nudging, behavioral nudges are seen as tools that unconsciously sway people's choices and behaviors (Thaler & Sunstein 2008). Nudges are not form of coercion: though they typically work by bypassing our full blown rational abilities they also do not completely force us to make a certain decision (Saghai 2013, 48). In contrast, cognitive nudges or nudges to reason "are designed to be processed by filters that are partially constitutive of reasoning in normal functioning agents, not an obstacle to reasoning" (Levy, 2017, p. 498). As such they help us exercise our autonomy of reasoning and they are not paternalistic tools.

When it comes to content moderation, this type of nudging comes in different shapes and forms but its explicit goal is to both curb lazy and emotional thinking and alert users of the epistemic norms of social media interaction. Here are some concrete examples:

Studies indicate that preventive debunking (or 'inoculation') makes users less subject to accepting misinformation. As it is explained in a recent study on misinformation and climate change, "inoculating messages that (1) explain the flawed argumentation technique used in the misinformation or that (2) highlight the scientific consensus on climate change, were effective in neutralizing the adverse effects of misinformation" (Cook et al 2017). This strategy can be adopted to incentivize users to care about truth , giving them the tools to be more reflective in analyzing what they see online. If this could be somehow embedded in the social media experience, it would be a powerful way to improve the epistemic environment.

Another type of nudging consists of shifting the focus on truth and accuracy. Here is a recent study on this: "participants were subtly primed to think about accuracy by being asked to rate the accuracy of a single news headline. This […] was sufficient to more than double participants' level of discernment between sharing true versus false headlines" (Pennycook et al 2020, p XX). Shifting the focus to truth allows participants to be alert and adopt their reflective stance and rationality rather than defaulting to their emotional or lazy thinking.

One last example comes from the world of social media itself. Recently Twitter has adopted a form of nudging that signals that epistemic norms are important in social media engagement. This strategy may be able to curb the gamification effect that, according to Nguyen (2021), is now predominant online. In this nudging effort, Twitter pushes users to be more careful when sharing content they have not read:

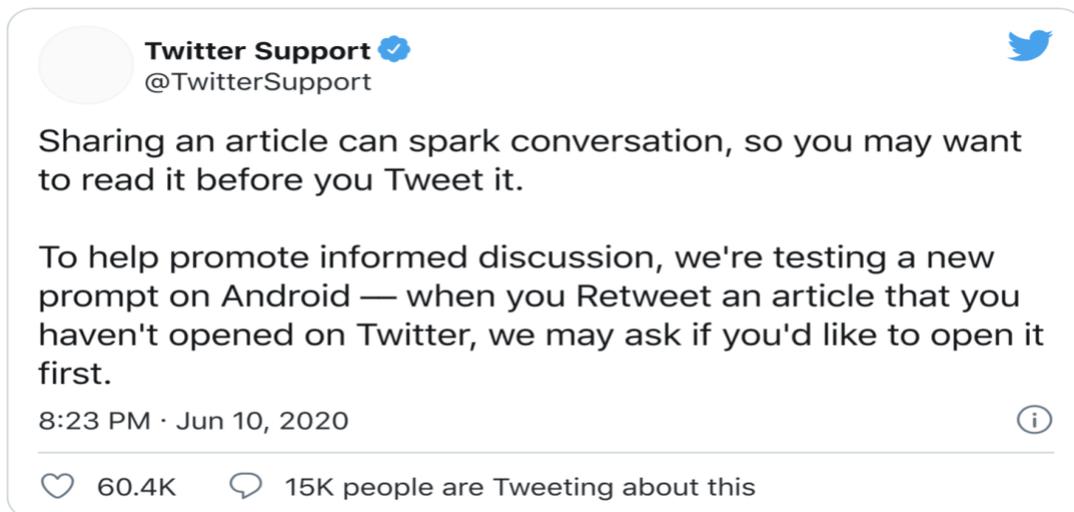

Some headlines may be catchy and spark a lot of interest, yet Twitter here is prompting users to take responsibility for what they share and its epistemic effects. This creates a 'friction' effect that slows them down and hopefully allows rationality to take over.

In conclusion, the envisaged solution here is to use the tools that foster our critical thinking skills making us more reflexive reasoners as we deal with online engagement. This is a form of light content moderation as it deals with structural problems on the platform without adopting any form of coercion. Also, this solution is non-paternalist through and through: it develops a form of nudging that *promotes* our autonomy and does not bypass our deliberative capacities while also avoiding all the problems of the more stringent content moderation tools. Indeed, this type of nudging does not constrain users' decisions but encourages them to become more reflective in their thinking. As such, these tools should be widely adopted by social media platforms at least as one of the ways to address misinformation not because necessarily harmful but because it is a symptom of a structural problem.